\documentclass[twocolumn]{jpsj2} 
\usepackage{bm}

\title{Ferromagnetic Quantum Critical Fluctuations and Anomalous Coexistence of Ferromagnetism and Superconductivity in UCoGe Revealed by Co-NMR and NQR Studies }

\author{
Tetsuya \textsc{Ohta,}$^{1}$\thanks{E-mail address: t-ohta@scphys.kyoto-u.ac.jp}
Yusuke \textsc{Nakai,}$^{1}$
Yoshihiko \textsc{Ihara,}$^{1}$
Kenji \textsc{Ishida,}$^{1}$\thanks{E-mail address: kishida@scphys.kyoto-u.ac.jp}
Kazuhiko \textsc{Deguchi,}$^{2}$
Noriaki K. \textsc{Sato}$^{2}$
and
Isamu \textsc{Satoh}$^{3}$
}

\inst{$^{1}$Department of Physics, Graduate School of Science, Kyoto University, Kyoto 606-8502, Japan. \\
$^{2}$Department of Physics, Graduate School of Science, Nagoya University, Nagoya 464-8602, Japan.\\
$^{3}$Institute for Materials Research, Tohoku University, Sendai 980-8577, Japan.
}

\abst{Co nuclear magnetic resonance (NMR) and nuclear quadrupole resonance (NQR) studies were carried out for the recently discovered UCoGe, in which the ferromagnetic and superconducting (SC) transitions are reported to occur at $T_{\rm Curie} \sim 3$ K and $T_S \sim 0.8$ K (N.~T.~Huy {\it et al.}, Phys. Rev. Lett. {\bf 99} (2007) 067006), in order to investigate the coexistence of ferromagnetism and superconductivity as well as the normal-state and SC properties from a microscopic point of view. From the nuclear spin-lattice relaxation rate $1/T_1$ and Knight-shift measurements, we confirm that ferromagnetic fluctuations that possess a quantum critical character are present above $T_{\rm Curie}$ and also the occurrence of a ferromagnetic transition at 2.5 K in our polycrystalline sample. The magnetic fluctuations in the normal state show that UCoGe is an itinerant ferromagnet similar to ZrZn$_2$ and YCo$_2$. The onset SC transition is identified at $T_S \sim 0.7$ K, below which $1/T_1$ arising from 30\% of the volume fraction starts to decrease due to the opening of the SC gap. This component of $1/T_1$, which follows a $T^3$ dependence in the temperature range $0.3 - 0.1$ K, coexists with the magnetic components of $1/T_1$ showing a $\sqrt{T}$ dependence below $T_S$. From the NQR measurements in the SC state, we suggest that the self-induced vortex state is realized in UCoGe.}

\kword{UCoGe, ferromagnetic superconductor, Co-NMR and NQR, coexistence}

\begin{document}
\maketitle

Since the discovery of superconductivity in ferromagnetic compounds under high pressure\cite{SaxenaNature00,AkazawaJPhys04}, the concept of the interplay between magnetism and superconductivity was changed, because ferromagnetism and superconductivity are considered to be mutually exclusive. The pressure studies in UGe$_2$ have shown that the superconducting (SC) transition temperature $T_S$ is higher in the pressure region where the ordered moments are enhanced, and that superconductivity is not observed when the ferromagnetism disappears in a further higher pressure region\cite{HuxleyPRB01}. These results suggest that ferromagnetism seems to enhance the superconductivity in UGe$_2$. In addition, ambient-pressure ferromagnetic superconductivity was discovered in URhGe, where $T_{\rm Curie} = 9.5$ K and $T_S \sim 0.25$ K were reported\cite{AokiNature01}. The relation between the ferromagnetism and superconductivity observed in the U compounds is one of the attractive topics for the community studying strongly correlated electron systems.
 
Quite recently, new ambient-pressure ferromagnetic superconductivity with $T_{\rm Curie} = 3$ K and $T_S \sim 0.8$ K was found in UCoGe\cite{HuyPRL07}. One of the advantages of UCoGe is that it includes ``NMR-active'' nucleus Co. In the NMR studies of ferromagnetic superconductors, Ge in UGe$_2$ is replaced by $^{73}$Ge with a nuclear spin and $^{73}$Ge nuclear quadrupole resonance (NQR) studies have been carried out by Kotegawa {\it et al.}\cite{KotegawaJPSJ05} and Harada {\it et al.}\cite{HaradaJPSJ05, HaradaPRB07}. They have reported the SC properties and the relation between ferromagnetism and superconductivity from the $^{73}$Ge-NQR results. In this paper, we show the detailed Co-NMR and NQR results of UCoGe, which are compared with the $^{73}$Ge-NQR results of high-pressure UGe$_2$. 
    
Polycrystalline UCoGe was prepared by the arc melting method. The details will be reported in a separate paper\cite{Sato}.  
\begin{figure}
\begin{center}
\includegraphics[width=7.2cm]{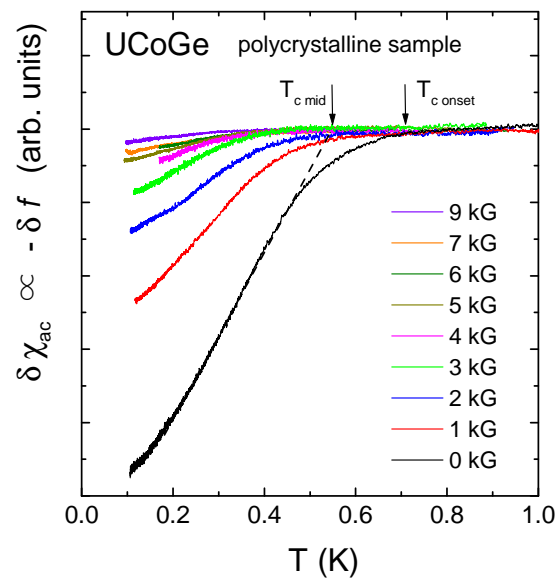}
\end{center}
\caption{(Color online) Temperature dependence of $\delta\chi_{\rm ac}$ under various magnetic fields below 9 kOe.}
\label{fig1}
\end{figure}
Figure 1 shows the temperature dependence of $\chi_{\rm ac}$ in various magnetic fields, which was obtained by measuring the self-inductance of an NMR coil with powdered UCoGe. A clear Meissner signal was observed in zero and magnetic fields, indicating that superconductivity persists up to 9 kOe. In zero field, the onset and midpoint of the SC transition are estimated to be 0.7 and 0.55 K, respectively. The Meissner behavior in zero field in our sample is consistent with that in a previous report\cite{HuyPRL07}.

\begin{figure}
\begin{center}
\includegraphics[width=7.5cm]{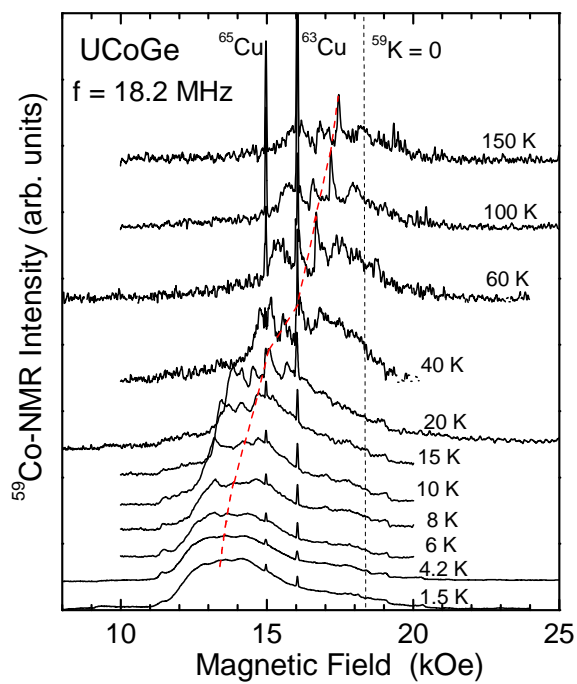}
\end{center}
\caption{(Color online) Field-sweep $^{59}$Co-NMR spectra obtained at 18.2 MHz at various temperatures. The sharp signals shown by $^{63}$Cu and $^{65}$Cu are the NMR signals arising from the NMR coil. $^{59}K = 0$ shows the field where the Knight shift of $^{59}$Co is zero. The dashed curve traces the NMR-signal peak arising from the $1/2 \leftrightarrow -1/2$ transition.  }
\label{fig2}
\end{figure}
Figure 2 shows the $^{59}$Co-NMR spectra of our powdered sample obtained by sweeping an external field at various temperatures. A symmetric NMR spectrum with several satellite peaks at higher temperatures, indicative of the presence of the electric field gradient (EFG) at the Co site, becomes anisotropic with decreasing temperature. The shift of the gravity of the broad NMR spectra originates from the isotropic component of the Knight shift. The NMR peak arising from the $1/2 \leftrightarrow -1/2$ transition shifts significantly with respect to temperature. The shift of this peak is traced by the dashed curve in Fig.~2.

\begin{figure}
\begin{center}
\includegraphics[width=7.5cm]{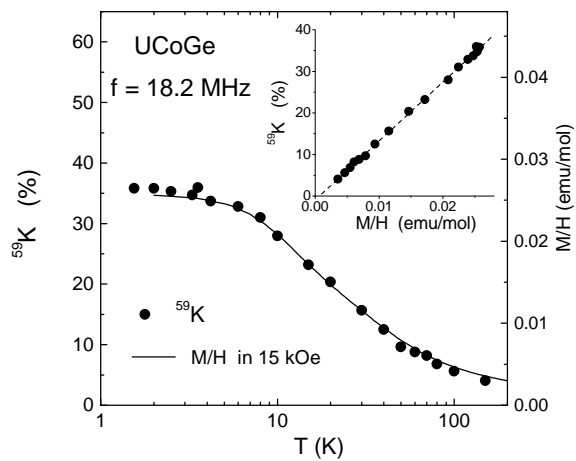}
\end{center}
\caption{Temperature dependence of the $^{59}$Co Knight shift evaluated from the $1/2 \leftrightarrow -1/2$ transition of the spectra shown in Fig.~2. The temperature dependence of $M /H$ measured in 15 kOe is shown in the same figure. The inset is the $^{59}K-M / H$ plot.}
\label{fig3}
\end{figure}
Figure 3 shows the temperature dependence of the Knight shift $^{59}K$ evaluated from the peak arising from the $1/2 \leftrightarrow -1/2$ transition, along with the magnetization $M$ divided by $H$ at 15 kOe. The temperature dependence of $^{59}K$ is well scaled with $M / H$. With decreasing $T$, $^{59}K$ becomes so large that it increases up to 35\% at low temperatures. It is confirmed that the Curie behavior in $M / H$ is microscopically intrinsic. It should be noted that the $^{59}K$ and Co-NMR spectra do not change through $T_{\rm Curie} \sim 3$ K, showing that the ferromagnetic transition is smeared out in the magnetic fields. This is characteristic of the ferromagnetic transition. From the linear relationship between $^{59}K$ and $M / H$, the hyperfine coupling constant $H_{\rm hf}$ at the Co site and the orbital shift $K_{\rm orb}$ are estimated to be $79 \pm 1$ kOe/$\mu_B$ and $-0.96 \pm 0.22 \%$, respectively. From the positive value of $H_{\rm hf}$ and the predominance of the isotropic term in the Knight shift, it is considered that the core polarization effect by Co-$3d$ electrons is not dominant but the transferred contribution from U-$5f$ electrons by way of Co-$4s$ orbits is dominant at the Co-nuclear site. The Co Knight-shift results indicate that U-$5f$ electrons play a crucial role for the ferromagnetism in UCoGe.             
   
\begin{figure}
\begin{center}
\includegraphics[width=7.5cm]{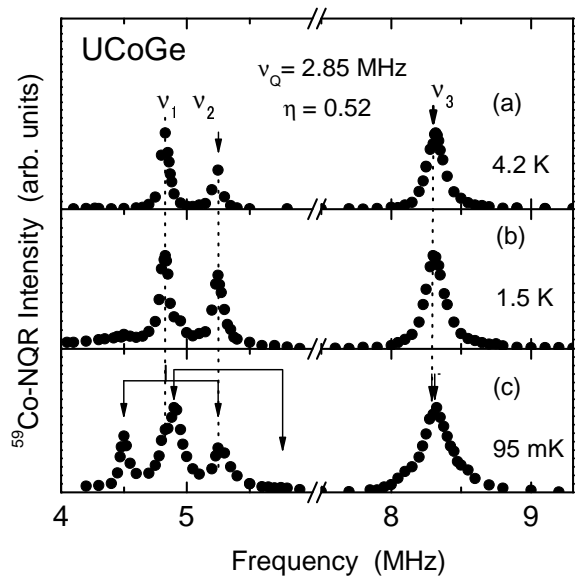}
\end{center}
\caption{
Co-NQR spectra at 4.2 K (above $T_{\rm Curie}$: (a)), 1.5 K (below $T_{\rm Curie}$: (b)), and 95 mK (the lowest temperature in the measurements: (c)). The NQR spectrum at 95 mK can be understood by the superposition of the two NQR spectra arising from the regions with and without the internal field. (Refer to the text). }
\label{fig1}
\end{figure}
Three Co-NQR signals were observed at the peaks of 4.8, 5.2, and 8.3 MHz at 4.2 K, as shown in Fig.~4. From these signals, we evaluated the NQR parameters $\nu_{zz}$, which is the NQR frequency along the principal axis of the EFG, to be 2.85 MHz and an asymmetric parameter $\eta$ to be 0.52. The large $\eta$ value shows that the symmetry of the Co site is lower than an axial symmetry. The Co-NQR spectra affected by the internal fields at the Co site were observed at low temperatures. The spectrum change by temperature is discussed below.  

\begin{figure}
\begin{center}
\includegraphics[width=8cm]{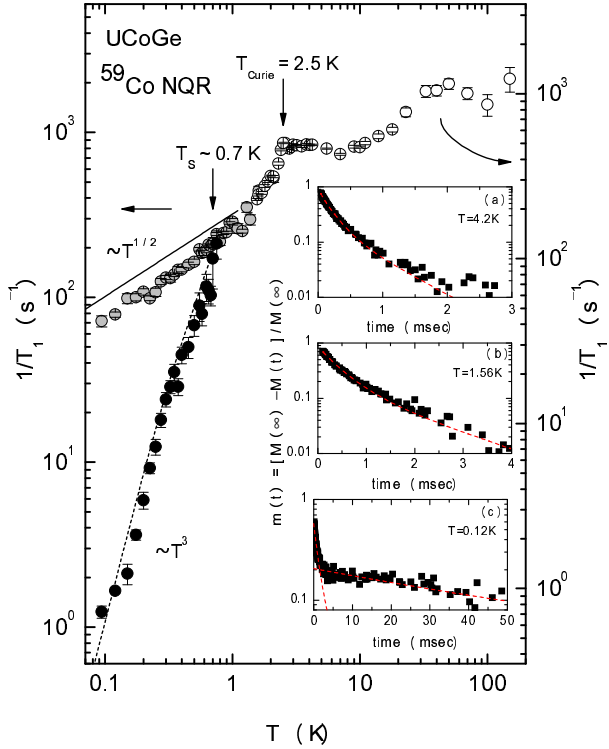}
\end{center}
\caption{(Color online) The main graph is the temperature dependence of $1/T_1$ measured at the Co-NQR signal of 8.3 MHz. The values of $1/T_1$ above (below) 1.5 K are shown by open (gray and black) circles with respect to the right (left) axis due to the difference in the $1/T_1$ values by the measurement conditions (refer to the text). The scale of the left-hand-side axis is 0.58 that of the right-hand-side axis. The insets are the recoveries of the nuclear magnetization after a saturation pulse. The dotted curves in the insets are the fitting curves for evaluating $1/T_1$. The recoveries above $T_{S}$ are fitted by a single component of $1/T_1$, as shown in the insets of (a) and (b). Two components of $1/T_1$ are observed below $T_{S}$, as shown in the inset of (c). The longer (black circles) and shorter (gray circles) components of $1/T_1$ are plotted in the main figure. The dotted curve below $T_S$ represents the temperature dependence of $1/T_1$ calculated using the 3-D polar model with $2\Delta_0/k_BT_S = 4$.    
}
\label{fig1}
\end{figure}
The nuclear spin-lattice relaxation rate 1/$T_1$ of Co in zero magnetic field was measured at the NQR signal observed at 8.3 MHz denoted as $\nu_3$ in Fig.~4.  Figure 5 shows the temperature dependence of $1/T_1$ of Co over the wide temperature range between 95 mK and 150 K. It was found that the values of $1/T_1$ below 10 K depend on the NQR pulse conditions. This is considered to be due to the development of the short-range ferromagnetic correlations. The details of this will be reported elsewhere\cite{Ohta}. Thus, the values of $1/T_1$ above and below 1.5 K are shown by different axes because the measurement conditions change around 1.5 K. The insets show the recovery curves $m(t)$ of the nuclear magnetization $M(t)$ at a time $t$ after a saturation pulse at three characteristic temperatures. Above $T_S$, the recovery curves were almost consistently determined with a single component of $1/T_1$, as observed in the insets (a) and (b).

First of all, we discuss the magnetic properties in the normal state on the basis of the $1/T_1$ and $^{59}K$ results. In general, $1/T_1$ divided by temperature $1/T_1T$ is related to the low-energy part of the $q$-dependent dynamical susceptibility in compounds. The spin-fluctuation character in compounds can be known from the comparison between $1/T_1T$ and $\chi$ or Knight shift.  
\begin{figure}
\begin{center}
\includegraphics[width=7.5cm]{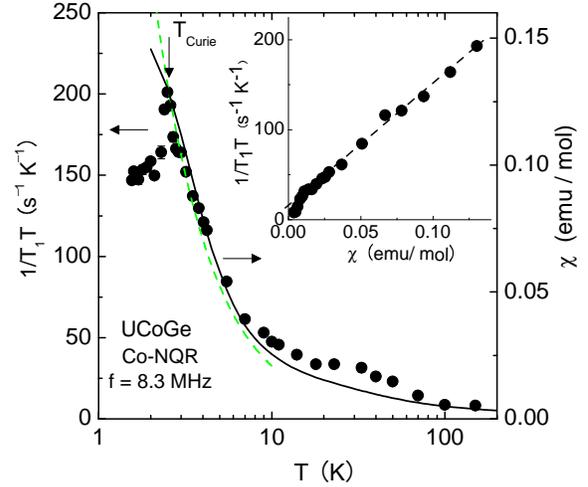}
\end{center}
\caption{(Color online) Temperature dependence of $1/T_1T$ in zero field and $\chi_{\rm bulk}$ measured in 1 kOe. The dashed curve shows a $T^{-4/3}$ dependence, which is expected at the FM critical point\cite{IshigakiJPSJ96}.  The inset is the plot of $1/T_1T$ against $\chi_{\rm bulk}$.}
\label{fig1}
\end{figure}
The main plot of Fig.~6 shows the temperature dependence of $1/T_1T$ in zero field, together with that of $\chi_{\rm bulk}$ in 1 kOe. The good linear relation was found between $1/T_1T$ and $\chi_{\rm bulk}$, as shown in the inset. This clearly proves that ferromagnetic (FM) fluctuations are dominant in UCoGe, since the dynamical susceptibility revealed by $1/T_1T$ is determined by the $q$ = 0 susceptibility. The relation $1/T_1T \propto \chi_{\rm bulk}$ is often observed in weak and nearly FM compounds\cite{KontaniSSC76,TakigawaJPSJ82,YoshimuraJPSJ84}, and it is interpreted by the SCR theory\cite{MoriyaJPSJ73P639,MoriyaJPSJ73P669}. Furthermore, we point out that the temperature dependence of $1/T_1T$ is close to a $T^{-4/3}$ dependence below 10 K, which is an expected behavior when a 3-D metal is close to the FM instability\cite{IshigakiJPSJ96}. The characteristic energy of the spin fluctuation $T_0$ is estimated to be $T_0 \sim 220$ K from the Sommerfeld coefficient $\gamma$\cite{HuyPRL07,MoriyaJPSJ95}, and according to the SCR calculation, $1/T_1T$ is close to a $T^{-4/3}$ dependence below $T / T_0 < 0.1$\cite{IshigakiJPSJ96}. Therefore, the observation of $1/T_1T \sim T^{-4/3}$ below 10 K is reasonable and suggests that UCoGe possesses an FM quantum critical character.
 
Although a clear anomaly was observed in $1/T_1T$ at $T_{\rm Curie}$, the NQR spectra of $\nu_1$ and $\nu_2$ are almost unchanged even at 1.5 K, as observed in Fig.~4, indicative of the absence of the static internal field down to 1.5 K. This might be related to the FM quantum critical character because it is anticipated that ferromagnetism develops gradually near the quantum critical point. By further lowering the temperature, new NQR signals affected by the static fields start to develop below 1 K. The Co-NQR spectra at 95 mK shown in Fig.~4 (c) can be understood with the superposition of two NQR spectra; one is the Co-NQR spectra without the internal field (``nonmagnetic'' NQR signals), and the other is the Co-NQR spectra affected by the internal field at the Co site (``magnetic'' NQR signal). The magnitude of the internal fields of the magnetic NQR signal is approximately 400 Oe, and its direction is almost perpendicular to the principal axis of the EFG. In Fig.~4, the calculated resonance frequencies are shown by arrow\cite{cal}. The NQR spectra in the FM state indicate that there exist two Co sites with and without the internal magnetic fields. Furthermore, the temperature variation of the NQR spectrum in the FM state shows a first-order-like behavior because the internal field of the magnetic NQR signal is independent of temperature and only a fraction of the magnetic and nonmagnetic NQR signals changes by due to temperature.      

Next, we discuss the SC properties revealed by the $1/T_1$ measurement. Below 0.7 K, which is the onset temperature of the SC transition, the longer component of $T_1$ ascribed to the opening of the SC gap was observed in $m(t)$, as shown in the inset (c) of Fig.~5. The longer and shorter components of $1/T_1$ are shown in the main panel of Fig.~5. The longer component of $1/T_1$ is estimated to be 30\% of the total relaxation component, which is determined by the extrapolation of $m(t)$ to $t = 0$, and it follows a $T^3$ dependence below 0.3 K down to 0.1 K. The $T^3$ dependence of $1/T_1$ suggests the presence of line nodes in the SC gap, and the complete $T$ dependence of $1/T_1$ below $T_S$ can be interpreted by the 3-D polar state with $\Delta(\theta) = \Delta_0\cos{\theta}$. The $T$ dependence of $1/T_1$ calculated using the model is shown in Fig.~5 with the dotted curve below $T_S$, where the magnitude of the SC gap is taken as $2\Delta_0/k_BT_S = 4$. The magnitude of the SC gap is quite comparable to that in UGe$_2$\cite{KotegawaJPSJ05,HaradaPRB07}. It should be noted that an appreciable Korringa ($T_1T$ = constant) behavior was not observed down to 95 mK, indicative of the small effect of the residual density of state. In contrast, the shorter component of $1/T_1$ shows a $\sqrt{T}$ dependence below $T_S$, suggestive of the persistence of the magnetic fluctuations down to 95 mK. It is quite unusual that magnetic fluctuations continue to develop in the SC state.

\begin{figure}
\begin{center}
\includegraphics[width=7.5cm]{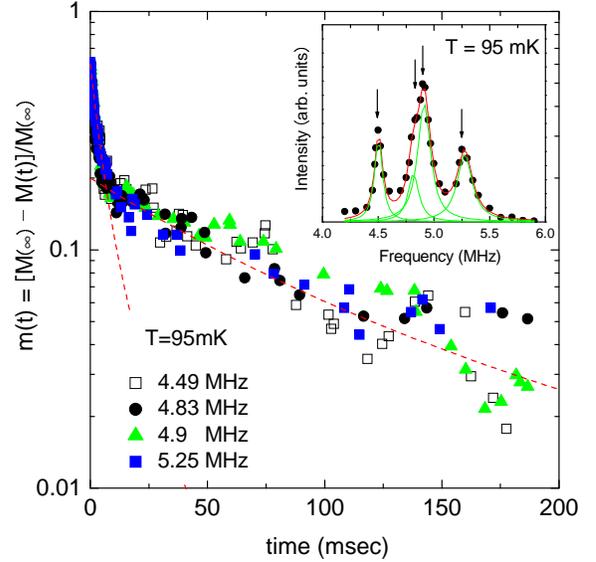}
\end{center}
\caption{(Color online) Co nuclear recoveries $m(t)$ measured at various frequencies at 95 mK.  The inset shows the NQR-signal peaks at 95 mK, where the recoveries are measured.}
\label{fig1}
\end{figure}
The underlying problem is the relation between the magnetism and superconductivity in UCoGe, i.e, whether superconductivity occurs only in the region without the internal field. To clarify this, we measure $1/T_1$ at 95 mK at various low-frequency NQR peaks. Figure 7 shows the recovery curves $m(t)$ measured at different frequencies. Here, the peaks at 4.83 and 5.25 MHz are the Co-NQR signals mainly arising from the region without the internal field, and the peaks at 4.49 and 4.9 MHz are the NQR signals appearing below $T_{\rm Curie}$, which are affected by the internal field, i.e. Co-NQR signals arising from the magnetic region. As observed in Fig.~7, $m(t)$ measured at the 4 different peaks shows the same behavior, indicating that the superconducting fractions (30\% of the total relaxation component) of the regions with and without the internal fields are not different. The same behavior of $m(t)$ at the 4 peaks excludes the possibility that superconductivity occurs in one of the regions with and without the internal field; rather, superconductivity occurs in both regions from the NQR spectra. This situation is different from the high-pressure ferromagnetic phase in UGe$_2$, where superconductivity is observed only in the ferromagnetic region\cite{HaradaJPSJ05}. We also point out that the short $1/T_1$ component is quite similar in both regions with and without the internal field. This is considered to be because the ordered moment is as small as 0.03$\mu_B$ at $T = 0$ K.    

Next, we attempt to understand the presence of the two components of $1/T_1$ in the entire sample. We suggest the possibility of the self-induced vortex (SIV) state, which is considered to be realized when superconductivity occurs in the ferromagnetic state\cite{TachikiSSC80}. The condition for the realization of the SIV state is $H_{c1} < 4\pi M < H_{c2}$, where $M$ is the net magnetization of a ferromagnet. When this condition is satisfied, the vortices are generated spontaneously. In this case, it is considered that $1/T_1$ in the SC state has two components arising from the SIV and SC regions, respectively. Recently, we observed two well-separated components of $1/T_1$ in the region close to the SC vortices in LaRu$_4$P$_{12}$\cite{NakaiPRL}. 

Another interesting possibility to be expected in UCoGe is the nonunitary spin-triplet state similar to the A1 state in $^3$He superfluidity, where the SC gap opens only in the up-spin band and not in the down-spin band\cite{OhmiPRL93}. In this case, it is considered that the Korringa behavior with 1/4 of the magnitude of $1/T_1T$ in the normal state is anticipated to remain at low temperatures, which arises from the relaxation process occurring in the down-spin intraband. This possibility has been discussed in UGe$_2$ from the $^{73}$Ge-NQR studies\cite{HaradaPRB07}. However, the absence of such Korringa behavior in the low-temperature region seems to exclude this possibility.

In conclusion, ferromagnetic and SC transitions were identified at $T_{\rm Curie} \sim 2.5$ K and $T_S \sim 0.7$ K in our polycrystalline sample. Ferromagnetic fluctuations that possess a quantum critical character are dominant above $T_{\rm Curie}$ from the $1/T_1$ and Knight-shift measurements. We found that $1/T_1$ equal to 30\% of the total relaxation component starts to decrease below $T_S$, accompanied by a $T^3$ dependence below 0.3 K, and that $1/T_1$ of the remaining component shows a $\sqrt{T}$ dependence, indicative of magnetic fluctuations even below $T_S$.  From the present NQR measurements in the SC state, we suggest that the self-induced vortex state is realized in UCoGe. 

We thank Y.~Maeno and K.~Yoshimura for experimental support and valuable discussions. 
We also thank N.~Tateiwa, H.~Kotegawa, Y.~Itoh, C.~Michioka, H.~Ikeda, and S. Fujimoto for valuable discussions.
This work was partially supported by the 21st COE program on ``Center for Diversity and Universality in Physics'' by MEXT of Japan, and ``Skutterudite'' and ``Super Clean'' by MEXT (Nos.1827006, 17071007) and Grants-in-Aid for Scientific Research from the Japan Society for the Promotion of Science (JSPS)(No.18340102).


\begin{thebibliography}{99}
\bibitem{cal}
For the calculation of the NQR frequencies affected by the internal magnetic field, the effect of the internal field is treated as perturbation. When the field was assumed to be $H = (H_0\sin\theta\cos\phi, H_0\sin\theta\sin\phi, H_0\cos\theta)$, the frequencies shown by the arrows are obtained by parameters of $(H_0,\theta,\phi)$ = (400 Oe, 85$^{\circ}$, 45$^{\circ}$).
\end{thebibliography}
\end{document}